\documentclass[prl,aps,twocolumn,nofootinbib,preprintnumbers]{revtex4}
\usepackage{graphicx}
\usepackage{epsfig,amssymb,amsmath,color}

\def\bea{\begin{eqnarray}}
\def\eea{\end{eqnarray}}

\begin{document}

\draft
\tighten
\preprint{CTPU-15-06, KIAS-P15021}
\title{\large \bf
Diluting the inflationary axion fluctuation by a stronger QCD in the early Universe
}
\author{
Kiwoon Choi$^{\,a}$\footnote{e-mail: kchoi@ibs.re.kr},
Eung Jin Chun$^{b}$\footnote{e-mail: ejchun@kias.re.kr},
Sang Hui Im$^{a}$\footnote{e-mail: shim@ibs.re.kr},
Kwang Sik Jeong$^{\,c}$\footnote{e-mail: ksjeong@pusan.ac.kr}
 }
\affiliation{$^a$ Center for Theoretical Physics of the Universe,
     IBS, Daejeon 305-811, Korea
     \\
     $^b$ Korea Institute for Advanced Study, Seoul 130-722, Korea
     \\
     $^c$ Department of Physics, Pusan National University, Busan, 609-735, Korea
    }

\vspace{2cm}

\begin{abstract}

We propose a new mechanism to suppress the axion isocurvature perturbation,
while producing the right amount of axion dark matter, within the framework of
supersymmetric axion models with the axion scale induced by supersymmetry breaking.
The mechanism involves an intermediate phase transition to generate the Higgs
$\mu$-parameter, before which the weak scale is comparable to the axion scale
and the resulting stronger QCD yields an axion mass heavier than the Hubble scale
over a certain period.
Combined with that the Hubble-induced axion scale during the primordial inflation
is well above the intermediate axion scale at present, the stronger QCD in the early
Universe suppresses the axion fluctuation to be small enough even when the inflationary
Hubble scale saturates the current upper bound, while generating an axion
misalignment angle of order unity.

\end{abstract}

\maketitle

The non-observation of the neutron EDM requires the CP violating QCD angle  to be as tiny
as $|\bar\theta| < 10^{-10}$, causing the strong CP problem.
An appealing solution of this puzzle is to introduce a spontaneously broken
global Peccei-Quinn (PQ) symmetry \cite{Peccei-Quinn}.
Then $\bar\theta$ corresponds to the vacuum value of the associated Nambu-Goldstone boson,
the axion, which is determined to be vanishing by the low energy QCD dynamics
\cite{axion-review}.

An interesting consequence of this solution is that axions can explain the dark matter
in our universe.
Yet, the prospect for axion dark matter depends on the cosmological history of the PQ
phase transition.
A possible scenario is that the spontaneous PQ breaking occurs after the primordial
inflation is over.
In such a case, the model is constrained to have the domain-wall number
$N_{\rm DW} = 1$, where $N_{\rm DW}$ corresponds to the integer-valued
U$(1)_{\rm PQ}\times$SU$(3)_c\times$SU$(3)_c$ anomaly coefficient.
Then axions are produced mainly by the annihilations of axionic strings and domain-walls,
which would result in the right amount of axion dark matter for the axion scale
$f_a \sim 5\times 10^{10}$~GeV \cite{axion-from-topological-defects}.
However it appears to be difficult to realize this scenario within the framework
of a fundamental theory such as string theory, since
it requires a PQ symmetry with $N_{\rm DW}=1$, as well as a restored PQ phase until
some moment after the primordial inflation.

Another scenario which we will focus on in this paper is that U$(1)_{\rm PQ}$
is spontaneously broken during the primordial inflation and never restored afterwards.
Then the model is not subject to the condition $N_{\rm DW}=1$, but is constrained
by the axion isocurvature perturbation
\cite{axion-cosmology1,axion-cosmology2,axion-cosmology3}.
For instance, from the observed CMB power spectrum, one finds
\cite{Planck-results},
\bea
\label{iso-constraint}
\left(\frac{\delta T}{T}\right)_{\rm iso}\,\simeq\,
\frac{4}{5}\left(\frac{\Omega_a}{\Omega_{\rm DM}}\right)
\frac{\delta\theta}{\theta_{\rm mis}}
\,<\, 3.8\times 10^{-6},
\eea
where $\theta_{\rm mis}$ and $\delta\theta$ denote the average misalignment angle
and the angle fluctuation, respectively, for the axion field right before
the conventional QCD phase transition when $m_a(t_{\rm QCD})\approx H(t_{\rm QCD})$
with a temperature $T(t_{\rm QCD})\sim 1$~GeV.
The relic axion density  is given by
\bea
\label{axion-density}
\frac{\Omega_a}{\Omega_{\rm DM}}
\simeq
1.7 \,\theta^2_{\rm mis}
\left( \frac{f_a(t_0)}{10^{12}\, {\rm GeV} }\right)^{1.19},
\eea
with $\Omega_{\rm DM}\approx 0.24$ being the total dark matter fraction.
Here we have assumed that
$|\delta\theta| \ll |\theta_{\rm mis}|$ and there is no significant evolution of $f_a$
from $t_{\rm QCD}$ to the present time $t_0$ so that $f_a(t_{\rm QCD})\approx f_a(t_0)$.
In inflationary cosmology, the primordial quantum fluctuation of the axion field
results in
\bea
\label{angle-fluctuation}
\delta \theta \,\equiv\, \delta\theta(t_{\rm QCD}) \,=\,
\gamma \delta\theta(t_I)\,=\, \gamma\frac{H(t_I)}{2\pi f_a(t_I)},
\eea
where $f_a(t_I)$ and $H(t_I)$ denote the axion scale and the Hubble parameter, respectively,
during {the primordial inflation epoch} $t_I$, and
the factor $\gamma$ is introduced to take into account the evolution of $\delta\theta$ from
$t_I$ to $t_{\rm QCD}$.
Note that the inflationary Hubble scale $H(t_I)$ is bounded by the tensor-to-scalar ratio
of the CMB perturbation as
\bea
\label{hubble-bound}
r \,\simeq\, 0.16\left(\frac{H(t_I)}{10^{14}{\rm GeV}}\right)^2 \, <\, 0.11,\eea
and the weak gravity conjecture \cite{weak-gravity-conjecture} suggests that
generic axion scales are bounded
as
\bea\label{weak-gravity}
f_a \,\lesssim\, {\cal O}\left(\frac{g^2}{8\pi^2}M_{Pl}\right),\eea
where $M_{Pl}\simeq 2.4\times 10^{18}$~GeV is the reduced Planck mass.

To discuss the implication of the isocurvature constraint (\ref{iso-constraint}),
one needs to specify the cosmological evolution of the axion scale after the primordial
inflation is over.
If $f_a(t_I)\sim f_a(t_0)$ as has been assumed in most of the previous studies,
it requires that either $H(t_I)$ is smaller than its upper bound $\sim 10^{14}$~GeV by
at least five orders of magnitude, so that the CMB tensor mode is too small
to be observable, {or} $\delta\theta$ should experience a large suppression
after the primordial inflation, which appears to be difficult to be implemented.

The above observation suggests a more attractive scenario realizing
$f_a(t_I)\gg f_a(t_0)$ \cite{Linde:1991km} in a natural manner.
Indeed supersymmetric axion models offer a natural scheme to realize such
a scenario, generating the axion scale through the competition between the {tachyonic}
SUSY breaking mass term and a supersymmetric, but Planck-scale-suppressed higher
dimensional term in the scalar potential \cite{fa-susy,Chun:2014xva,Choi:2014uaa,chun04}.
One then finds
\bea
\label{axion-scale}
f_a(t_0) &\sim& \sqrt{m_{\rm SUSY}M_{Pl}},
\nonumber
\\
f_a(t_I) &\sim & \sqrt{H(t_I)M_{Pl}},
\eea
which explains elegantly the origin of an intermediate axion scale at present, while
giving a Hubble-induced inflationary axion scale well above  the present axion scale,
if the supersymmetry (SUSY) breaking mass $m_{\rm SUSY}$ at present is around TeV scale.
Furthermore, this type of axion models can be successfully embedded into string theory.
Specifically, they can be identified as a low energy limit of string models involving
an anomalous U$(1)_A$ gauge symmetry with vanishing Fayet-Illiopoulos term
\cite{Choi:2014uaa,axion-U(1)A}.
In such string models, the U$(1)_A$ gauge boson is decoupled from the low energy world
by receiving a heavy mass $M_A\sim g^2 M_{Pl}/8\pi^2$ through the St\"uckelberg mechanism,
while leaving the global part of U$(1)_A$ as an unbroken PQ symmetry
in the supersymmetric limit.
Once SUSY breaking is introduced properly, in both the present Universe and
the inflationary early Universe, the residual PQ symmetry can be spontaneously broken
to generate the axion scales as (\ref{axion-scale}).

In this paper, we discuss a novel mechanism to suppress the axion isocurvature perturbation,
while producing the right amount of axion dark mater,
within the framework of supersymmetric axion models with the axion scales given
by (\ref{axion-scale}).
The isocurvature constraint (\ref{iso-constraint}) and the relic axion density
(\ref{axion-density}) suggest that for $H(t_I)$ near the current upper bound $\sim 10^{14}$~GeV,
the allowed amount of axion dark matter is maximal when $f_a(t_0)\sim 10^{11}$--$10^{13}$~GeV,
while $f_a(t_I)$ nearly saturates the weak gravity bound (\ref{weak-gravity}),
e.g.~$f_a(t_I)\sim 10^{16}$--$10^{17}$~GeV.
Interestingly, the axions scales generated by SUSY breaking as (\ref{axion-scale}) automatically
realize such pattern if $m_{\rm SUSY}$ is around TeV scale.
More specifically, for the case
\bea
f_a(t_0)/f_a(t_I)\approx \sqrt{m_{\rm SUSY}/H(t_I)},
\eea
the isocurvature bound (\ref{iso-constraint}) reads off
\bea
\frac{H(t_I)}{10^{14} {\rm GeV}} \,<\,
 \left(\frac{0.08}{\gamma}\right)^2
\left(\frac{\Omega_{\rm DM}}{\Omega_a}\right)
\left(\frac{f_a(t_0)}{10^{12}{\rm GeV}}\right)^{0.8}
\left(\frac{1{\rm TeV}}{m_{\rm SUSY}}\right),
\nonumber
\eea
when combined with (\ref{axion-density}).
This implies that a high scale inflation scenario with
$H(t_I)\sim 10^{13}$--$10^{14}$~GeV, which would give an observable tensor-to-scalar ratio
$r={\cal O}(0.1$--$0.01)$ in the CMB perturbation, can be compatible with the axion dark matter
$\Omega_a=\Omega_{\rm DM}$, if the axion field fluctuation experiences just a mild suppression
after $t_I$, e.g.~$\gamma={\cal O}(0.1$--$0.01)$ in (\ref{angle-fluctuation}).

To suppress $\delta\theta$ through its cosmological evolution, one needs a period with
$m_a(t) > H(t)$ well before $t_{\rm QCD}$.
On the other hand, usually this is not easy to be realized because the axion mass should
be generated mostly by the QCD anomaly in order for the strong CP problem solved
by the PQ mechanism.
(See Refs.~\cite{Jeong:2013xta,Higaki:2014ooa,Fairbairn:2014zta,Kitajima:2014xla}
for an alternative possibility.)
In the following, we propose a simple scheme to achieve such a cosmological period
by having a phase of stronger QCD in the early Universe.

Our scheme is based on a phase transition at $t=t_\mu \gg t_I$, which will be called
the $\mu$-transition in the following as it generates the Higgs $\mu$-parameter through
the superpotential term \cite{Kim-Nilles-mechanism},
\bea
\label{kim-nilles}
\mu(X)H_uH_d \,\equiv\, \frac{\kappa_1 X^2 H_uH_d}{M_{Pl}},
\eea
where $X$ is a PQ-charged gauge-singlet superfield.
Specifically,
\bea
X(t\leq t_\mu)=0, \quad
X(t>t_\mu)\sim \sqrt{m_{\rm SUSY}M_{Pl}},
\eea
so that
\bea
\mu(t\leq t_{\mu})=0, \quad
\mu(t>t_{\mu}) \sim m_{\rm SUSY}.
\eea
With this transition, the weak scale experiences an unusual evolution in a way
that the weak scale before the $\mu$-transition is comparable to the axion
scale (\ref{axion-scale}), as will be discussed below.

To proceed, let us discuss first the key features of the scheme, and later present
an explicit model to realize the whole ingredients.
Including the Hubble-induced contribution, the mass of the $D$-flat Higgs direction $H_uH_d$
is generically given by
\bea
m_{\phi}^2 = c_\phi H^2 + \xi_\phi m_{\rm SUSY}^2 + 2|\mu|^2 \quad (\phi^2\equiv H_uH_d),
\eea
where $c_\phi$ and $\xi_\phi$ are model-dependent parameters of order unity.
In our scheme, both $c_\phi$ and $\xi_\phi$ are assumed to be negative, so $m_{\phi}^2<0$
before the $\mu$-transition.
Then $\phi=\sqrt{H_uH_d}$ is stabilized by the competition between the tachyonic
$m^2_{\phi}|\phi|^2$ and a supersymmetric term of ${\cal O}(|\phi|^6/M_{Pl}^2)$
in the scalar potential, which results in
\bea
f_a(t_I)&\sim & \phi(t_I)\,\sim\, \sqrt{H(t_I)M_{Pl}},
\nonumber \\
f_a(t_\mu)&\sim & \phi(t_\mu)\, \sim\, \sqrt{m_{\rm SUSY}M_{Pl}}.
\eea
On the other hand, after the $\mu$-transition, $m_{\phi}^2>0$ due to $\mu\sim m_{\rm SUSY}$.
The resulting weak scale and axion scale at present are given by
\bea
\phi(t_0) &= & {\cal O}(100) \,\, {\rm GeV} , \nonumber \\
f_a(t_0) &\sim &   X(t_0)\, \sim\,  \sqrt{m_{\rm SUSY}M_{Pl}}.
\eea

A simple consequence of the above evolution of $H_uH_d$ is that the weak scale is comparable
to the axion scale before the $\mu$-transition:
\bea
\tilde \phi \equiv\, \phi({t\leq t_\mu})\,\sim\, \tilde f_a \,\equiv\, f_a(t\leq t_\mu).
\eea
This results in a higher QCD scale, i.e. a stronger QCD, and therefore a heavier axion mass
which might be even bigger than the Hubble scale for a certain period.
Let us estimate the QCD scale $\tilde \Lambda_{\rm QCD}$ before the $\mu$-transition,
which is defined as the scale where the 1-loop QCD coupling blows up, as well as the resulting
axion mass $\tilde m_a$.
For the case with
$\tilde \Lambda_{\rm QCD} <\tilde m_{\tilde g}(\tilde m_{\tilde g}) < 10^{-5} \tilde \phi$,
where $\tilde m_{\tilde g}$ denotes the gluino mass before the $\mu$-transition,
we find
\bea
\tilde \Lambda_{\rm QCD}&
\approx & 23\,{\rm TeV}
\left(\frac{\tilde m_{\tilde g}}{30 \,{\rm TeV} }\right)^{2/11}
\nonumber \\
&&
\times
\left(\frac{\tan\beta}{10}\right)^{3/11}
\left(\frac{\tilde \phi}{10^{12}\,{\rm GeV}}\right)^{6/11},
\eea
where $\tan\beta=\langle H_u \rangle/\langle H_d \rangle$ at present,
and $\tilde m_{\tilde g}/\tilde g_3^2(\tilde m_{\tilde g}) \simeq
m_{\tilde g}/{g^2_3(m_{\tilde g})}$ for the gluino mass $m_{\tilde g}$ at present.
Here we assume that $g_3^2(M_{\rm GUT})=\tilde g_3^2(M_{\rm GUT})$ and
$y_q(M_{\rm GUT})=\tilde y_q(M_{\rm GUT})$ for the QCD coupling and the quark Yukawa couplings.
When the temperature $T\lesssim \tilde\Lambda_{\rm QCD}$,
the axion mass during the period of stronger QCD is estimated to be
\bea
\label{axion-mass-1}
\tilde m_a \,\approx\, 
\tilde \Lambda^{2}_{\rm QCD}/{\tilde f_a}.
\eea
On the other hand, if $\tilde m_{\tilde g}(\tilde \Lambda_{\rm np})
< \tilde\Lambda_{\rm QCD} < 10^{-5} \tilde \phi$,
the resulting QCD scale is estimated as
\bea
\tilde \Lambda_{\rm QCD}
&\approx&
21\, {\rm TeV}
\left(\frac{\tan\beta}{10}\right)^{1/3}
\left(\frac{\tilde \phi}{10^{12}{\rm GeV}}\right)^{2/3},
\eea
with the axion mass
\bea
\label{axion-mass-2}
\tilde m_a \,\approx\,
\tilde m^{1/2}_{\tilde g} \tilde \Lambda^{3/2}_{\rm QCD}/{\tilde f_a}.
\eea
Here $\tilde \Lambda_{\rm np}$ denotes the scale where the stronger QCD becomes
nonperturbative, i.e.~around $\tilde g^2_3=8\pi^2/N_c$ with $N_c=3$.
Note that the axion potential for the axion mass (\ref{axion-mass-2}) can be
obtained by a single insertion of the SUSY breaking spurion $\tilde m_{\tilde g}\theta^2$
to the nonperturbative superpotential $W_{\rm np}\sim \tilde \Lambda_{\rm QCD}^3$
induced by the gluino condensation.

If the stronger QCD scale $\tilde\Lambda_{\rm QCD}$ is high enough,
there could be a period with $\tilde m_a(t) > H(t)$ well before the conventional
QCD phase transition.
As is well known, in such a period the axion field experiences a damped oscillation,
with an amplitude $\bar a$ (averaged over each oscillation period) evolving  as
\bea
\bar a \, \propto R^{-3/2}(t),
\eea
where $R(t)$ is the scale factor of the expanding universe.
Then the spatially averaged vacuum value of the axion field is settled down at the minimum
of the axion potential induced by the stronger QCD, while the axion angle fluctuation
is diluted according to
\bea
\label{angle-fluc}
\delta \theta \,=\, \gamma \frac{H(t_I)}{2\pi f_a(t_I)} \,\approx\,
\left(\frac{T(t_\mu)}{T(t_i)}\right)^{3/2}\frac{H(t_I)}{2\pi f_a(t_I)},
\eea
where $t=t_i$  denotes the moment when the damped axion oscillation begins, and $t=t_\mu$ is
the moment when it is over.
Note that, after the $\mu$-transition, the weak scale and the QCD scale quickly roll down
to the present values, so the axion mass becomes negligible compared to $H(t)$ until
$t\sim t_{\rm QCD}$ when the Universe undergoes the conventional QCD phase transition.
Also, the minimum of the axion potential induced by the stronger QCD  is generically
different from the minimum of the axion potential at present.
As a result, our scheme generates an axion misalignment angle of order unity:
\bea
\theta_{\rm mis}\,\equiv\, \left\langle \frac{a(t_\mu)}{f_a(t_\mu)}\right\rangle
- \left\langle \frac{a(t_0)}{f_a(t_0)}\right\rangle \,=\, {\cal O}(1),
\eea
together with an intermediate axion scale at present, so gives rise to
$\Omega_a=\Omega_{\rm DM}$ in a natural way.

In our case, the damped axion oscillation induced by the stronger QCD begins at a temperature
$T(t_i)\sim \tilde\Lambda_{\rm QCD}$ as $\tilde m_a$ is highly suppressed by thermal effects
for $T\gg \tilde\Lambda_{\rm QCD}$.
On the other hand, the scalar field $X$ generating $\mu$ through
(\ref{kim-nilles}) is trapped at the origin by thermal effects until the Universe cools
down to a temperature $T(t_\mu)\sim m_{\rm SUSY}$.
In fact, our scheme involves a variety of dimensionless parameters which affect the naive
estimate of the involved scales.
We find that there is a large fraction of the natural parameter region where
the axion mass
\bea
\tilde m_a \,\approx\, 0.4\, {\rm MeV}
\left(\frac{\tilde f_a}{10^{12}{\rm GeV}}\right)^{-1}
\left(\frac{\tilde \Lambda_{\rm QCD}}{20{\rm TeV}}\right)^2
\eea
is larger than the Hubble scale
\bea
\label{hubble}
H(t_\mu) \simeq 0.2\, {\rm MeV}\left(\frac{\sqrt{V_0}}{1{\rm TeV}\times 10^{12}{\rm GeV}}\right),
\eea
over the period $t_i\lesssim t\lesssim t_\mu$ with a temperature ratio:
\bea
T(t_\mu)/T(t_i)= {\cal O}(10^{-1}\mbox{--}10^{-2}).
\eea
Then the resulting $\delta\theta$ given by (\ref{angle-fluc}) can be small enough to satisfy
the isocurvature bound (\ref{iso-constraint}) even when $H(t_I)$ saturates its upper bound
$\sim 10^{14}$~GeV.
Note that during $t_i\lesssim t\lesssim t_\mu$,
\bea
\phi(t)-\phi(t_0)\,\sim\, X(t)-X(t_0)\, \sim\, \sqrt{m_{\rm SUSY}M_{Pl}},
\nonumber
\eea
so the corresponding vacuum energy density $V_0={\cal O}(m^2_{\rm SUSY} f_a^2(t_0))$.
This means that in this period the Universe is dominated by the vacuum energy density
with the Hubble scale given by (\ref{hubble}), which is often called the thermal inflation
\cite{thermal-inflation}.

\begin{figure}[t]
  \begin{center}
  \begin{tabular}{l}
   \includegraphics[trim=0.4cm 0.5cm -0.4cm 0, width=0.5\textwidth]{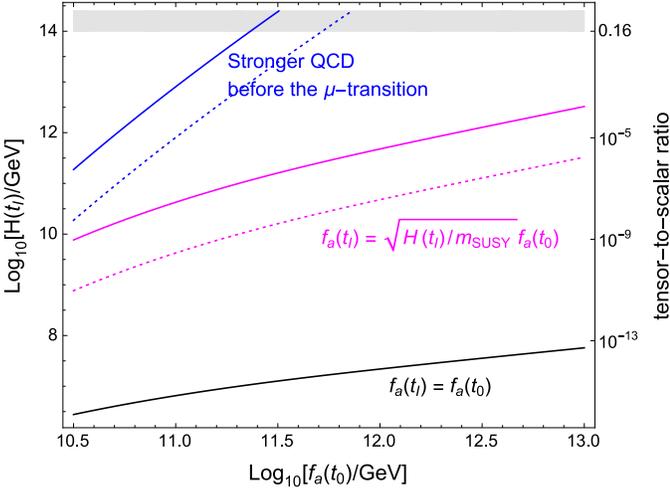}
   \end{tabular}
  \end{center}
  \caption{
  Upper bound on the inflationary Hubble scale consistent with the axion dark matter,
  $\Omega_a=\Omega_{\rm DM}$.
  Here we have taken $m_{\tilde g}=3$~TeV, $\tan\beta=10$, and $T(t_\mu)=1$~TeV.
  The shaded region is excluded by the Planck results. The black solid line
  is the constraint in the conventional scenario with $f_a(t_I)=f_a(t_0)$.
  The magenta lines are for the scenario with $f_a(t_I)/f_a(t_0) = \sqrt{H(t_I)/m_{\rm SUSY}}$,
  but without a stronger QCD.
  The blue lines are for our scheme which leads to a further suppression of $\delta\theta$
  by the stronger QCD.
  The SUSY breaking mass has been taken $m_{\rm SUSY}=1$~TeV for the solid lines
  and 10~TeV for the dotted lines.
  }
\label{fig:iso-bound}
\end{figure}

It should be stressed that in our scheme the axion isocurvature perturbation is
suppressed by two steps.
The first suppression is due to $f_a(t_0)/f_a(t_I)\sim \sqrt{m_{\rm SUSY}/{H(t_I)}}\ll 1$,
and the second is due to the stronger QCD dynamics before the $\mu$-transition, yielding
a further suppression by $\gamma \sim (m_{\rm SUSY}/\tilde\Lambda_{\rm QCD})^{3/2}$.
To illustrate the result, we depict in Fig.~\ref{fig:iso-bound} the upper bound on the inflationary
Hubble scale $H(t_I)$ resulting from the isocurvature constraint (\ref{iso-constraint})
for $\Omega_a=\Omega_{\rm DM}$.
To make a comparison, we depict the results for three distinct cases:
$(i)$ the conventional scenario of $f_a(t_I)=f_a(t_0)$ without a stronger QCD,
$(ii)$ a scheme with $f_a(t_I)/f_a(t_0)\sim \sqrt{H(t_I)/m_{\rm SUSY}}$, but without
a stronger QCD, $(iii)$ our scheme with  $f_a(t_I)/f_a(t_0)\sim \sqrt{H(t_I)/m_{\rm SUSY}}$
and a stronger QCD before the $\mu$-transition.

Let us now present an explicit model implementing the mechanisms discussed above.
As a simple example, we consider a model with the following superpotential,
\bea
\label{model-1}
&&W \,=\, (\mbox{MSSM Yukawa terms}) + \lambda Y\Phi\Phi^c
\nonumber \\
&& + \,\frac{\kappa_1 X^2H_uH_d}{M_{Pl}}
+ \frac{\kappa_2XY^3}{M_{Pl}}+\frac{\kappa_3(H_uH_d)(LH_u)}{M_{Pl}}, \quad\quad
\eea
where $X$ and $Y$ are PQ-charged gauge singlets responsible for the $\mu$-transition,
$L$ is the MSSM lepton doublet, and $\Phi+\Phi^c$ are $U(1)_Y$-charged exotic matter fields
introduced to give a thermal mass to $Y$.
Then the scalar potential for the $\mu$-transition is given by
\bea
\label{mu-transition}
V_1 &=& m_X^2|X|^2+m_Y^2|Y|^2 +\left(\frac{\kappa_2A_2}{M_{Pl}}XY^3+{\rm h.c.}\right)
\nonumber \\
&&
+\,\frac{|\kappa_2|^2}{M_{Pl}^2}\left(|Y|^6+9|X|^2|Y|^4\right),
\eea
where
\bea
m_X^2 &=&c_XH^2+\xi_Xm_{\rm SUSY}^2 + 4|\mu_X|^2,
\nonumber \\
m_Y^2 &=& c_YH^2+\xi_Ym_{\rm SUSY}^2+ \alpha_YT^2,
\eea
for $\mu_X = \kappa_1 H_uH_d/M_{Pl}$.
Here $c_{X,Y}H^2$ are the Hubble-induced masses, $\xi_{X,Y} m_{\rm SUSY}^2$ are
the SUSY breaking masses at zero temperature, and
$\alpha_Y T^2$ is the thermal mass of $Y$ induced by the coupling $\lambda Y\Phi\Phi^c$,
which is of ${\cal O}(|\lambda|^2T^2)$ for $|\lambda Y|< T$.

For simplicity, we will assume that all the dimensionless parameters
appearing in the superpotential and the SUSY breaking scalar masses are of order unity.
However it should be noted that these parameters can have a variation
of ${\cal O}(0.1$--$10)$ easily. In particular, the superpotential parameters $\kappa_n$
can have a much wider variation without invoking fine-tuning.
This gives us a rather large room to get an enough suppression
of the axion angle fluctuation $\delta\theta$ through a stronger QCD before the
$\mu$-transition.
At any rate, assuming that $c_{X,Y}>0$, $\xi_X> 0$
and $\xi_Y<0$, the scalar potential (\ref{mu-transition}) indeed yields
the desired $\mu$-transition as
\bea
\label{vev1}
&& X=Y=0 \quad {\rm at}\,\,\, t\leq t_\mu,
\nonumber \\
&& X\sim Y \sim \sqrt{m_{\rm SUSY}M_{Pl}} \quad {\rm at} \,\,\, t>t_\mu,
\eea
with $T(t_\mu)\sim m_{\rm SUSY}$.

Now the Higgs and slepton fields can have a nontrivial evolution along the following
flat direction:
\bea
&& H^T_d = (\phi_d,0),\quad
L^T = (\phi_\ell,0),
\nonumber \\
&& H^T_u = (0,\sqrt{|\phi_d|^2+|\phi_\ell|^2} ),
\eea
which satisfies the $F$ and $D$ flat conditions.
The relevant terms of the scalar potential of $\phi_{d,\ell}$ are given by
\bea
V_2&=&
\sum m_i^2|\phi_i|^2 + \sqrt{\sum |\phi_i|^2}
\Big(B\mu\phi_d + {\rm h.c.} \Big)
\nonumber \\
&+&
\left(\sum |\phi_i|^2\right) \left(\frac{\kappa_3 A_3\phi_d\phi_\ell}{M_{Pl}}
+ {\rm h.c.}\right)
\nonumber \\
&+&
\frac{|\kappa_3|^2}{M_{Pl}^2}\left(\sum |\phi_i|^2\right)\left(
|\phi_d|^4+4|\phi_d\phi_\ell|^2+|\phi_\ell|^4 \right),
\quad\quad
\eea
for $\mu=\kappa_1X^2/M_{Pl}$, where
\bea
m_{\phi_d}^2 &=& c_d H^2 +\xi_d m_{\rm SUSY}^2 + 2|\mu|^2,
\nonumber\\
m_{\phi_\ell}^2&=& c_\ell H^2 +\xi_\ell m_{\rm SUSY}^2 +|\mu|^2.
\eea
Again, assuming $c_{d,\ell}<0$ and $\xi_{d,\ell}<0$, but
$m_{\phi_d,\phi_\ell}^2(t_0) >0$ due to $\mu(t_0)\sim m_{\rm SUSY}$,
the above scalar potential yields
\bea
\label{vev2}
f_a(t_I)\sim \phi_{d,\ell}(t_I)\sim \sqrt{H(t_I)M_{Pl}},
\nonumber
\\
f_a(t_\mu)\sim \phi_{d,\ell}(t_\mu)\sim \sqrt{m_{\rm SUSY}M_{Pl}},
\eea
and
\bea
\label{vev3}
\phi_d(t_0) &=&{\cal O}(100) \,\, {\rm GeV}, \quad \phi_\ell(t_0) \,=\, 0,
\nonumber \\
f_a(t_0) &\sim& X(t_0) \,\sim\, Y(t_0)\,\sim\, \sqrt{m_{\rm SUSY}M_{Pl}}.
\eea

To summarize, under a reasonably plausible assumption on the SUSY breaking during the primordial
inflation and in the present Universe, the model with the superpotential (\ref{model-1})
can successfully realize the desired cosmological evolution of the three relevant scales:
the axion scale, the weak scale, and the QCD scale as given by (\ref{vev1}), (\ref{vev2}) and (\ref{vev3}).
Being generated by SUSY breaking,
an inflationary axion scale $f_a(t_I)\sim \sqrt{H(t_I)/m_{\rm SUSY}}\, f_a(t_0)$
is determined to be well above the present axion scale $f_a(t_0) \sim \sqrt{m_{\rm SUSY}M_{Pl}}$,
and a stronger QCD in the early
Universe is realized to yield an enough suppression of the axion angle fluctuation even when
$H(t_I)$ saturates its upper bound.
We note that the minimum of the axion potential induced by the stronger QCD depends
on $\arg(\kappa_3A_3)$, but not on $\arg(B\mu)$,
while the minimum of the axion potential at present depends on $\arg(B\mu)$, but not on $\arg(\kappa_3A_3)$.
As a result, the stronger QCD generates an axion misalignment angle
$\theta_{\rm mis}={\cal O}(1)$,
so that the axion dark matter with $\Omega_a=\Omega_{\rm DM}$ arises naturally in our scheme.

There is a remaining issue which should be addressed to complete our scheme.
As we have noticed, the $\mu$-transition is foregone by a late-time thermal inflation.
This suggests that the scheme should be accompanied by a late-time baryogenesis operating
after the $\mu$-transition.
In fact, the model of (\ref{model-1}) offers an elegant mechanism to generate the baryon
asymmetry through the rolling flat direction $LH_u$ \cite{AD-mechanism}.
More detailed cosmology of our scheme, including the leptogenesis by rolling $LH_u$,
will be discussed elsewhere \cite{in-preparation}.

\section*{Acknowledgment}

This work was supported by IBS under the project code, IBS-R018-D1~[KC and SHI],
and by Pusan National University Research Grant, 2015~[KSJ],
and the Research Fund Program of Research Institute for Basic Sciences,
Pusan National University, Korea, 2015, Project No.~RIBS-PNU-2015-303~[KSJ].

\end{document}